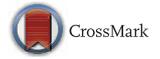

# Design of optimised backstepping controller for the synchronisation of chaotic Colpitts oscillator using shark smell algorithm

EHSAN FOULADI and HAMED MOJALLALI*

Electrical Engineering Department, Faculty of Engineering, University of Guilan, Rasht, Iran
*Corresponding author. E-mail: mojallali@guilan.ac.ir

**Abstract.** In this paper, an adaptive backstepping controller has been tuned to synchronise two chaotic Colpitts oscillators in a master–slave configuration. The parameters of the controller are determined using shark smell optimisation (SSO) algorithm. Numerical results are presented and compared with those of particle swarm optimisation (PSO) algorithm. Simulation results show better performance in terms of accuracy and convergence for the proposed optimised method compared to PSO optimised controller or any non-optimised backstepping controller.

## 1. Introduction

Colpitts oscillators are used in many fields such as radars, sensors, convertors, chaos generators, encryption, etc. [1–5]. Due to their wide range of applications, they have attracted more attention in recent years. The periodic output of Colpitts oscillator becomes chaotic under some unique circumstances. Therefore, a controller is required to control or synchronise chaotic signals.

Various methods such as sliding mode control method [6], PID controllers [7], adaptive control method [8], etc. are proposed for controlling chaotic oscillators. Backstopping control method is a nonlinear approach that recursively utilises compound Lyapunov function interlaced with feedback control. In this paper, adaptive backstepping control method is adopted and optimised using shark smell optimisation (SSO) algorithm to yield better results in comparison with conventional backstepping controllers.

Shark smell optimisation algorithm is inspired by sharks' smelling sensation and uses its hunting methods to find the optimal minima. It is faster-converging and avoids getting trapped in local minima [9,10]. Due to these features, SSO algorithm is suitably used to optimise the controller parameters for chaotic synchronisation. Furthermore, the simulation results are presented to confirm the robustness of the method.

The rest of the paper is organised as follows: in §2, Colpitts oscillators are introduced briefly and in §3, the adaptive backstepping method for chaos control and synchronisation is proposed. Section 4 discusses the shark smell optimisation algorithm and the utilisation procedure. In §5, the proposed cost function for the optimal controller is presented. Numerical results are presented in §6. Finally, conclusions are presented in §7.

## 2. Colpitts oscillators

Figure 1 shows the conventional Colpitts oscillator consisting of a bipolar junction transistor (BJT) along with an inductor and a pair of capacitors as the resonance network [11]. Dynamics of Colpitts oscillator are described using differential equations as follows:

$$\frac{dx}{dt} = y - aF(z)$$

$$\frac{dy}{dt} = c - x - by - z$$

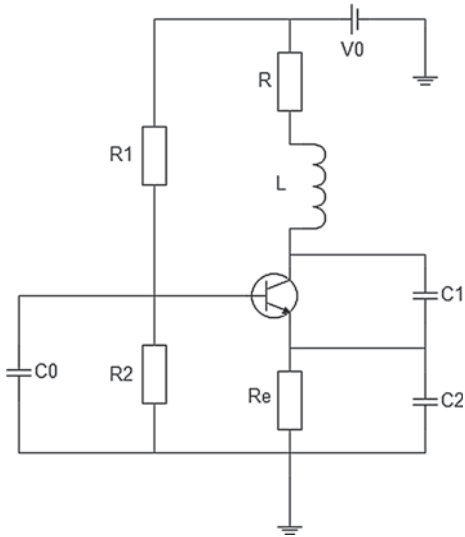

**Figure 1.** Conventional Colpitts oscillator.

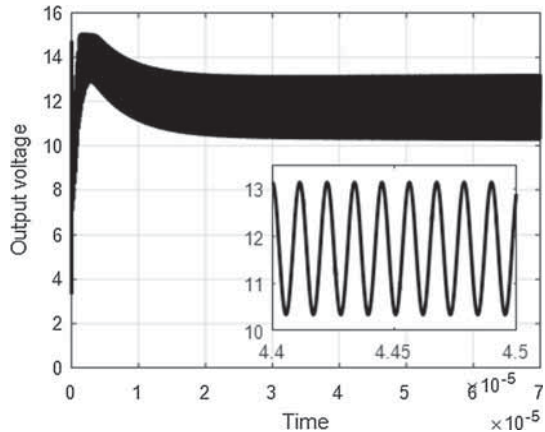

**Figure 2.** Periodic output signal of a generic Colpitts oscillator oscillating at 9 MHz.

$$\frac{dz}{dt} = y - dz, \quad (1)$$

in which $x = V_{C_1}/V^*$, $y = \rho I_L/V^*$, $z = V_{C_2}/V^*$, $\rho = \sqrt{LC_1}$, $a = \rho/r$, $b = R/\rho$, $c = V_0/V^*$, $d = \rho/R_e$, $e = [R_2/(R_1 + R_2)]c$,

$$F(x) = \begin{cases} e - 1 - z, & z < e - 1 \\ 0, & z \geq e - 1. \end{cases}$$

More detailed description on physical parameters and electrical characteristics of the oscillator is available in ref. [12]. Periodic output signal of a generic Colpitts oscillator is presented in figure 2. The circuit parameters are as follows: $C_1 = C_2 = 470$ nF, $C_0 = 47$ μF, $R = 36$ Ω, $R_e = 510$ Ω, $R_1 = R_2 = 3$ KΩ, $V_0 = 15$ V. The typical parameters of the Colpitts oscillator are $a = 30$, $b = 0.8$, $c = 20$, $d = 0.08$, $e = 10$. The phase waveform of the chaotic Colpitts oscillator is presented in figure 3.

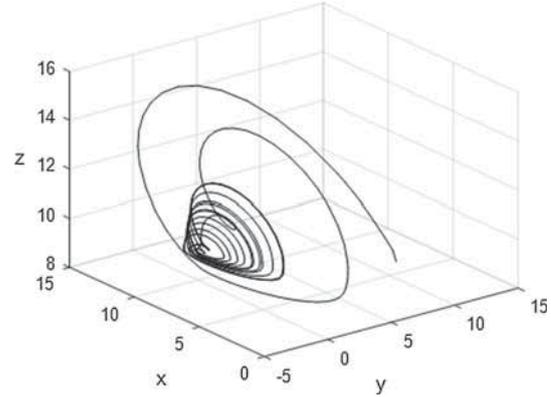

**Figure 3.** Phase waveform of the chaotic Colpitts oscillator.

## 3. Adaptive backstepping method

In this section, design of an appropriate controller is explained and applied to alter the chaotic signal in a desired manner [13]. The backstepping technique is a step-by-step design approach that utilises Lyapunov function [14] interlaced with feedback control at each step. At the last step, the main controller is acquired. The controlled chaotic system which is known as the slave system is as follows:

$$\frac{dx}{dt} = y - aF(z) + U$$

$$\frac{dy}{dt} = c - x - by - z$$

$$\frac{dz}{dt} = y - dz, \quad (2)$$

where $U$ is the controller. Let the ideal solution and the target orbit of the system be $X_t = (x_t, y_t, z_t)^T$, which satisfies the system (eq. (2)). So the master system is as follows:

$$\frac{dx_t}{dt} = y_t - aF(z_t)$$

$$\frac{dy_t}{dt} = c - x_t - by_t - z_t$$

$$\frac{dz_t}{dt} = y_t - dz_t. \quad (3)$$

So, the synchronisation of the slave system (eq. (2)), with the master system (eq. (3)) is applied by designing the appropriate controller. Error dynamic equations are obtained through subtraction of the systems mentioned, i.e., $\lim_{t\to\infty}|X_t - X| = 0$, and are presented as follows [15]:

$$\frac{de_1}{dt} = e_2 - aF(z_t) + aF(z_t - e_3) - U$$

$$\frac{de_2}{dt} = -e_1 - be_2 - e_3$$

$$\frac{de_3}{dt} = e_2 - de_3, \quad (4)$$

where the error vector is: $e = (e_1, e_2, e_3)^T = (x_t - x, y_t - y, z_t - z)^T$.

Now, the goal is to stabilise the variables of system (4) at the origin using the controller. By stabilising the system, the error vector $e$ reaches zero as time goes to infinity. The design procedure of the controller using the backstepping method is as follows:

*Step* 1: Stabilise the $e_3$-subsystem:

$$\dot{e}_3 = e_2 - de_3, \quad (5)$$

where $e_2 = -k_1(e_3)$, $k_1 \geq 0$ is considered as a virtual controller. Choosing the Lyapunov function $v_1 = \frac{1}{2}e_3^2$, its derivative is $\dot{v}_1 = -de_3^2 + e_3(-k_1(e_3))$. For a positive $k_1$, $\dot{v}_1$ is negative definite. So, the $e_3$-subsystem is asymptotically stable. Since the virtual controller $-k_1 e_3$ is estimated when $e_2$ is considered as a controller, the error between $e_2$ and $-k_1 e_3$ is $w_2 = e_2 + k_1 e_3$. So the $(e_3, w_2)$-subsystem has been obtained as follows:

$$\dot{e}_3 = w_2 - (k_1 + d)e_3$$

$$\dot{w}_2 = -e_1 + (k_1 - b)w_2$$
$$\quad + (bk_1 - 1 - k_1^2 - dk_1)e_3, \quad (6)$$

where $e_1 = -k_2(e_3, w_2)$, $k_2 \geq 0$ is considered as the virtual controller to stabilise the system asymptotically.

*Step* 2: The Lyapunov function $v_2 = v_1 + \frac{1}{2}w_2^2$ is chosen. Its derivative is $\dot{v}_2 = -(d + k_1)e_3^2 - (b - k_1)w_2^2 + w_2(-k_2(e_3, w_2))$. If $k_1 < b$, then $\dot{v}_2$ is negative definite. So, this makes the $(e_3, w_2)$-subsystem asymptotically stable. Likewise, assume $w_3 = e_1 + k_2(e_3, w_2)$, so that $\dot{w}_2 = -w_3 + (k_1 - k_2 - b)w_2 - e_3$. Here is the full dimension, $(e_3, w_2, w_3)$-subsystem presented below:

$$\dot{e}_3 = w_2 - (k_1 + d)e_3$$

$$\dot{w}_2 = -w_3 + (k_1 - k_2 - b)w_2 - e_3$$

$$\dot{w}_3 = -aF(z_t) + aF(z_t - e_3) - U + k_2 w_3$$
$$\quad + (k_1^2 - bk_1 + dk_1 - k_1 k_2 + k_2^2 + bk_2 + 1)w_2$$
$$\quad + (bdk_1 - k_1 + bk_1^2 - k_1^3 - 2dk_1^2 - d^2 k_1$$
$$\quad + k_2)e_3. \quad (7)$$

*Step* 3: Choose the Lyapunov function as $v_3 = v_2 + \frac{1}{2}w_3^2$, to make the $(e_3, w_2, w_3)$-subsystem stable. Its derivative is expressed as $\dot{v}_3 = -(d + k_1)e_3^2 - (b - k_1 + k_2)w_2^2 + k_2 w_3^2 + w_3(-k_3(e_3, w_2, w_3))$, $k_3 \geq 0$. If $k_2 \leq 0$, then $\dot{v}_3$ is negative definite. From Step 1, we have $k_2 \geq 0$. So, $k_2 = 0$. Let $U = -aF(z_t) + aF(z_t - e_3) + k_3 w_3 + (k_1^2 - bk_1 + dk_1)w_2 + (dk_1 - k_1 + bk_1^2 - k_1^3 - 2dk_1^2 - d^2 k_1)e_3$ and then $\dot{v}_3 < 0$. Therefore, the $(e_3, w_2, w_3)$-subsystem is asymptotically stable. Finally, the full $(e_3, w_2, w_3)$ system is as below:

$$\dot{e}_3 = w_2 - (k_1 + d)e_3$$

$$\dot{w}_2 = -w_3 + (k_1 - b)w_2 - e_3$$

$$\dot{w}_3 = -k_3 w_3 + w_2. \quad (8)$$

In the $(e_3, w_2, w_3)$-system coordinates, the equilibrium $(0, 0, 0)$ is globally asymptotically stable. Assuming $w_2 = e_2, w_3 = e_1$, equilibrium of (4) is still $(0, 0, 0)$. That is, the trajectory of the controlled system (2) asymptotically approaches the master signal (ideal solution $X_t$).

## 4. Shark smell optimisation algorithm

Shark smell optimisation (SSO) algorithm has been proposed by Abedinia *et al* [16]. This algorithm is based on sharks' distinct smelling ability for seeking the prey. In shark's hunting, concentration of the odour is an important factor. Likewise, the shark moves in the direction with higher odour density. This characteristic is used in SSO algorithm to find the solution of an optimisation problem. In this section, SSO algorithm has been explained briefly. For further information refer to [16].

A random population of vectors is generated within the feasible search domain as the initial position. Each position vector is a possible next step destination for the shark. The cost of each position indicates its nearness to the prey. By increasing the density of the odour particles, the position and velocity of the shark will be changed. The velocity model is based on the gradient of the objective function, tilting it towards the orientation with the lowest objective function. The velocity for the $m$th stage is represented as follows:

$$V_i^m = \mu_m \cdot R_1 \cdot \nabla(\text{OF})|_{X_i^m} + \alpha_m \cdot R_2 \cdot V_i^{m-1},$$

$$i = 1, \ldots, \text{NP}, \quad m = 1, \ldots, M, \quad (9)$$

where $\nabla(\text{OF})$ represents the gradient of objective function, $\mu_m$ is the gradient constant and $R_1, R_2 \in (0, 1)$ are random numbers with uniform distribution. $\alpha_m \in (0, 1)$ is the inertia constant for stage $m$. The velocity has an upper bound. Equation (10) shows the upper bound for a single dimension.

$$|V_{i,j}^m| = \text{Min}\left[\left|\mu_m \cdot R_1 \cdot \frac{\partial(\text{OF})}{\partial x_j}\bigg|_{X_{i,j}^m}\right.\right.$$
$$\left.\left. + |\alpha_m \cdot R_2 \cdot V_{i,j}^{m-1}|, |\gamma_m \cdot V_{i,j}^{m-1}|\right]\right.$$
$$i = 1, \ldots, \text{NP}, \quad j = 1, \ldots, \text{ND}, \quad m = 1, \ldots, M, \tag{10}$$

where $\gamma_m$ represents the upper bound of the current stage velocity. NP, ND and $M$ are the number of positions, the number of dimensions which in this paper is 2 and the number of stages, respectively. Next stage position is obtained as follows:

$$Y_i^{m+1} = X_i^m + V_i^m \cdot \Delta T_m$$
$$i = 1, \ldots, \text{NP}, \quad m = 1, \ldots, M, \tag{11}$$

where $\Delta T_m = 1$ is the time interval for the $m$th stage, that usually is considered one in all stages for simplicity.

$$Z_i^{m+1,k} = Y_i^{m+1} + R_3 \cdot Y_i^{m+1}$$
$$i = 1, \ldots, \text{NP}, \quad m = 1, \ldots, M, \quad k = 1, \ldots, K, \tag{12}$$

where $R_3 \in (-1, 1)$ is a random number and $K$ is the number of points for local search. Among the points searched in global and local scale, one with the lowest cost is selected to be the next destination:

$$X_i^{m+1} = \arg(\min(\text{OF}(Y_i^{m+1}), \text{OF}(Z_i^{m+1,1}),$$
$$\ldots, \text{OF}(Z_i^{m+1,K}))) \quad i = 1, \ldots, \text{NP}. \tag{13}$$

In the selection mechanism a minimisation problem is assumed, which is needed to minimise the error in chaos synchronisation.

## 5. Proposed optimal controller

Three parameters ($k_1, k_2, k_3$) are the positive parameters of the backstepping controller to be optimised among which $k_2 = 0$ and $k_1 < b$. In the present study, SSO algorithm is chosen to optimise the variable parameters. Error state variables ($e_i$) are used as the inputs in eq. (14), to calculate total sum of squares (TSS) which is used in optimisation algorithm as the cost function. The main goal is to minimise the cost function which is considered as follows:

$$\text{TSS} = \sum_{i=1}^{d} \int_0^{t_f} e_i^2(t) \, \mathrm{d}t. \tag{14}$$

TSS is the sum of the integral of squared errors in three dimensions. $t_f$ is the final time, $n$ is the number of controller parameters and $e_i(t)$ ($i = 1, 2, 3$) are the difference between the master and the slave systems. The goal of this paper is to stabilise the error system to zero. Error state variables converge to zero by minimising the fitness function using SSO algorithm.

## 6. Numerical results

In this section, optimisation algorithms are used to optimise the adaptive backstepping controller in order to synchronise two chaotic signals. Initial conditions for the slave and the master signals are respectively considered as $X_0 = (8, 2, 3)$, $X_{t_0} = (10.45, 0.718, 8.89)$. Results for the controller parameters' optimisation are presented in tables 1 and 2. All the simulations are done, under the same circumstances, 50 for population size through 30 number of iterations. A comparison between SSO and generic PSO algorithm is presented in figure 4.

Using the obtained optimal parameters from the SSO algorithm, a synchronisation between two identical Colpitts oscillators is acquired. In the first 20 s the controller is disabled. After $t = 20$ s, the controller is activated.

**Table 1.** Results of SSO algorithm.

| Algorithm | Experiment | $K_1$ | $K_3$ | TSS |
|---|---|---|---|---|
| SSO ($K_2 = 0$) | 1 | 2.8642e-05 | 2.4988 | 42.3754 |
| | 2 | 1.4077e-07 | 2.4982 | 42.3754 |
| | 3 | 1.0327e-06 | 2.4981 | 42.3754 |
| | 4 | 5.2964e-06 | 2.4916 | 42.3755 |
| | 5 | 4.8248e-06 | 2.4978 | 42.3754 |
| | 6 | 1.2834e-06 | 2.4975 | 42.3754 |
| | 7 | 1.1684e-06 | 2.496 | 42.3754 |
| | 8 | 1.4593e-05 | 2.4947 | 42.3754 |
| | 9 | 2.8652e-06 | 2.498 | 42.3754 |
| | 10 | 1.5512e-05 | 2.4974 | 42.3754 |

**Table 2.** Results of PSO algorithm.

| Algorithm | Experiment | $K_1$ | $K_3$ | TSS |
|---|---|---|---|---|
| PSO ($K_2 = 0$) | 1 | 0.096789 | 2.6962 | 42.5193 |
| | 2 | 0.025651 | 2.5226 | 42.4102 |
| | 3 | 0.025559 | 2.1146 | 42.4988 |
| | 4 | 0.010515 | 2.6926 | 42.4042 |
| | 5 | 0.016949 | 2.5962 | 42.4039 |
| | 6 | 0.011816 | 2.4675 | 42.396 |
| | 7 | 0.065559 | 2.5367 | 42.4675 |
| | 8 | 0.015606 | 2.446 | 42.4129 |
| | 9 | 0.29803 | 2.9554 | 42.9509 |
| | 10 | 0.26078 | 2.3763 | 42.9131 |

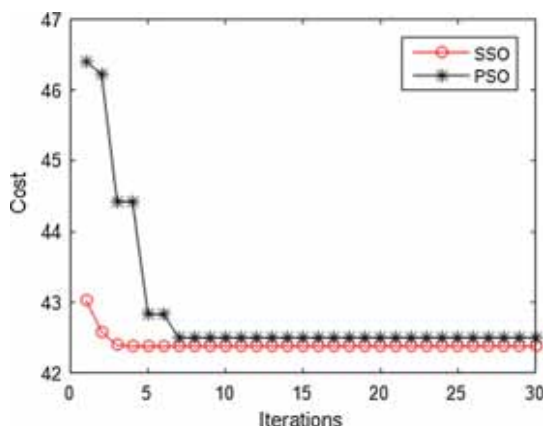

**Figure 4.** A comparison between SSO and generic PSO algorithm.

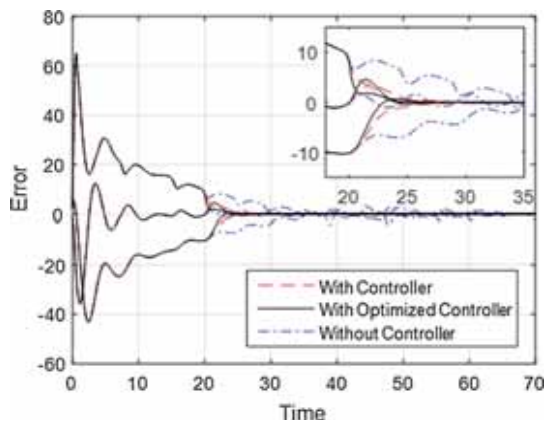

**Figure 5.** The waveform of the error state variables.

The simulation results are shown in figure 5. It shows the waveform of the error state variables. The simulations are done from $t = 0$ to $t = 70$ s.

As shown in figure 5, SSO is faster converging and the results are better, in terms of minimising the cost function. Best parameters using SSO algorithm are (1.4077e-07, 0, 2.4982), which has the cost of 42.4102.

## 7. Conclusion

In this paper, an optimal technique is demonstrated to synchronise Colpitts oscillator using optimised adaptive backstepping controller. The controller parameters are obtained by minimising the proposed cost function using the SSO algorithm. The results verify the effectiveness of the proposed optimised controller in achieving faster convergence on controlling chaos.

## References

[1] L Liu, R X Ma, H Xu, W K Wang, B J Wang and J X Li, Experimental investigation of a UWB direct chaotic through-wall imaging radar, *IET International Radar Conference 2015*
[2] K Bushida, K Mohri and T Uchiyama, *IEEE Trans. Mag.* **31**(**6**), 3134 (1995)
[3] Won Sup Chung and Kenzo Watanabe, *IEEE Trans. Instrum. Measure.* **34**(**4**), 534 (1985)
[4] L Jingxia and M Fuchang, Chaos generation in microwave band using improved Colpitts oscillator, in; *IEEE International Conference on Computing, Measurement, Control and Sensor Network (CMCSN)* (2012) pp. 405–408
[5] Yannick Abanda and Alain Tiedeu, *IET Image Processing* **10**(**10**), 742 (2016)
[6] Hao Zhang, Ma Xi-Kui and Liu Wei-Zeng, *Chaos Solitons Fractals* **21**(**5**), 1249 (2004)
[7] Wei-Der Chang and Yan Jun-Juh, *Chaos Solitons Fractals* **26**(**1**), 167 (2005)
[8] Shihua Chen and L Jinhu, *Chaos Solitons Fractals* **14**(**4**), 643 (2002)
[9] James Kennedy, Particle swarm optimization, in: *Encyclopedia of machine learning* (Springer, US, 2011) pp. 760–766
[10] Marco Dorigo, Birattari Mauro and Thomas Stutzle, *IEEE Comput. Intel. Mag.* **1**(**4**), 28 (2006)

[11] G M Maggio, Oscar De Feo and Michael Peter Kennedy, *IEEE Trans. Circuits Syst. I* **46**(**9**), 1118 (1999)

[12] Michael Peter Kennedy, *IEEE Trans. Circuits Syst. I* **41**(**11**), 771 (1994)

[13] Guo Hui Li, Shi Ping Zhou and Kui Yang, *Chaos Solitons Fractals* **33**(**2**), 582 (2007)

[14] Kazuo Tanaka, Tsuyoshi Hori and Hua O Wang, *IEEE Trans. Fuzzy Systems* **11**(**4**), 582 (2003)

[15] Antanas Cenys *et al*, *Chaos Solitons Fractals* **17**(**2**), 349 (2003)

[16] Oveis Abedinia, Nima Amjady and Ali Ghasemi, *Complexity* (2014)